# Quantum Software Development Lifecycle

Benjamin Weder, Johanna Barzen, Frank Leymann, and Daniel Vietz

**Abstract** With recent advances in the development of more powerful quantum computers, the research area of quantum software engineering is emerging, having the goal to provide concepts, principles, and guidelines to develop high-quality quantum applications. In classical software engineering, lifecycles are used to document the process of designing, implementing, maintaining, analyzing, and adapting software. Such lifecycles provide a common understanding of how to develop and operate an application, which is especially important due to the interdisciplinary nature of quantum computing. Since today's quantum applications are, in most cases, hybrid, consisting of quantum and classical programs, the lifecycle for quantum applications must involve the development of both kinds of programs. However, the existing lifecycles only target the development of quantum or classical programs in isolation. Additionally, the various programs must be orchestrated, e.g., using workflows. Thus, the development of quantum applications also incorporates the workflow lifecycle. In this chapter, we analyze the software artifacts usually comprising a quantum application and present their corresponding lifecycles. Furthermore, we identify the points of connection between the various lifecycles and integrate them into the overall quantum software development lifecycle. Therefore, the integrated lifecycle serves as a basis for the development and execution of hybrid quantum applications.

**Keywords:** Quantum Software Development, Quantum Computing, Hybrid Quantum Applications, Software Engineering, Software Lifecycle

Benjamin Weder
University of Stuttgart, Insitute of Architecture of Application Systems, Universitätsstraße 38, 70569 Stuttgart, Germany, e-mail: benjamin.weder@iaas.uni-stuttgart.de

Johanna Barzen
University of Stuttgart, Insitute of Architecture of Application Systems, Universitätsstraße 38, 70569 Stuttgart, Germany, e-mail: johanna.barzen@iaas.uni-stuttgart.de

Frank Leymann
University of Stuttgart, Insitute of Architecture of Application Systems, Universitätsstraße 38, 70569 Stuttgart, Germany, e-mail: frank.leymann@iaas.uni-stuttgart.de

Daniel Vietz
University of Stuttgart, Insitute of Architecture of Application Systems, Universitätsstraße 38, 70569 Stuttgart, Germany, e-mail: daniel.vietz@iaas.uni-stuttgart.de

# 1 Introduction

Quantum computing promises to solve many problems more efficiently or precisely than possible with classical computers, e.g., simulating complex physical systems or applying machine learning techniques [5, 6, 25]. With recent advances in developing more powerful quantum computers, also the development of corresponding quantum software and applications, as well as their integration into existing software architectures, are becoming increasingly important [49, 66]. However, the development of such quantum applications is complex and requires the knowledge of experts from various fields, e.g., physics, mathematics, and computer science [18, 60, 86].

*Quantum software engineering* is an emerging research area investigating concepts, principles, and guidelines to develop, maintain, and evolve quantum applications [66, 67, 97]. Thereby, it has the goal to increase the quality and reusability of the resulting quantum applications by systematically applying software engineering principles during all development phases from the initial requirement analysis to the retirement of the software [42, 86]. In classical software engineering, *software development lifecycles* are often used to document the different development phases a software artifact or application goes through [14, 59]. Furthermore, such software development lifecycles also summarize best practices and methods that can be applied in the various phases, as well as corresponding tools [86, 97]. Hence, they can be used for educating new developers by providing an overview of the development process or serve as a basis for the cooperation of experts from different fields [30].

Today's quantum applications are most often hybrid, consisting of quantum and classical programs [46, 65]. Thus, the lifecycle for quantum applications involves the development and operation of both kinds of programs. However, existing lifecycles from classical software engineering [14, 55], as well as quantum software lifecycles [18, 86], only target one of these kinds and do not address the resulting integration challenges. Furthermore, the execution of the quantum and classical programs must be orchestrated, and data has to be passed between them [90]. *Workflow technology* is a means for these orchestrations providing benefits, such as scalability, reliability, and robustness [20, 51]. Therefore, also the workflow lifecycle must be integrated into the overall lifecycle for developing quantum applications.

To address this, we introduce a *quantum software development lifecycle* describing the different relevant phases when developing and operating quantum applications. Thereby, we analyze the purpose of each phase, as well as available concepts and tools. Furthermore, we discuss the different software artifacts usually constituting a quantum application and present their corresponding lifecycles. Finally, we identify the plug points between the various lifecycles to enable their integration into our overall lifecycle for the development of hybrid quantum applications.

The remainder of this chapter is structured as follows: Section 2 introduces fundamentals about hybrid quantum applications. In Section 3, we present our quantum software development lifecycle, as well as the lifecycles of the different software artifacts constituting a hybrid quantum application. Afterwards, Section 4 describes assumptions and possible limitations of the introduced lifecycle. Finally, Section 5 discusses related work, and a conclusion and outlook are given in Section 6.

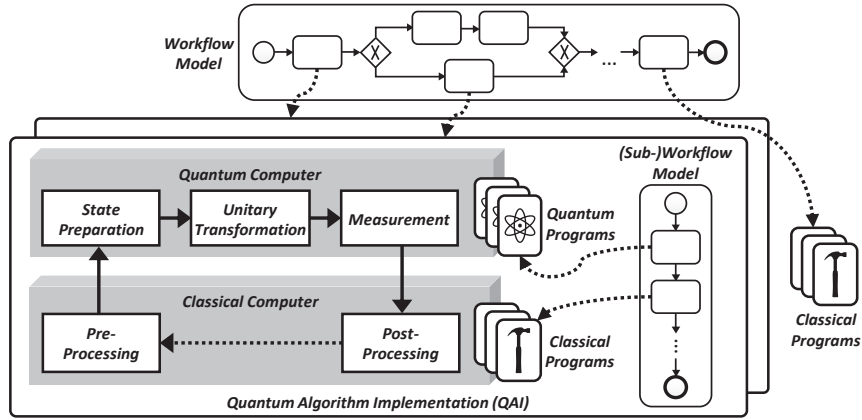

**Fig. 1** General Structure of a Hybrid Quantum Application.

## 2 Hybrid Quantum Applications

Nowadays, quantum applications are, in most cases, hybrid, i.e., they consist of *quantum algorithm implementations (QAIs)* and *classical programs*, as depicted in Figure 1 [46, 47, 76]. Thereby, the hybrid quantum application may comprise multiple QAIs, e.g., first performing clustering and then training a classifier based on the clustering results [5]. Furthermore, classical programs might be used to load data, transform it into another format, or visualize it for the user [47, 88].

But even a single QAI is often hybrid, comprising *quantum programs* and classical programs [46, 56]. The general structure of a gate-based QAI, i.e., quantum programs are realized as *quantum circuits*, is shown at the bottom of Figure 1 [46, 76]. Thereby, the *pre-processing* tasks are implemented by classical programs and executed on classical computers. Pre-processing, e.g., includes generating *state preparation* circuits based on input data to initialize the register of the quantum computer when executing the quantum programs [16, 91]. The quantum programs are executed on a quantum computer, first preparing the required state in the register depending on the generated state preparation circuit [46, 60]. Afterwards, the *unitary transformation* specified by the proper quantum algorithm is performed, and finally, the result is measured. *Post-processing*, e.g., on a classical computer, interprets the measurement results or mitigates readout-errors in the result distribution by applying an unfolding technique to retrieve a less disturbed distribution from the measured distribution [13, 54].

In addition, various quantum algorithms also require algorithm-specific pre- or post-processing steps that have to be executed on a classical computer [47, 56]. For example, the factorization algorithm of *Shor* [73] relies on classical post-processing to analyze continued fractions. Another example is *Simon's algorithm* [74], which requires solving a linear system of equations after the quantum computation. Further, different variational algorithms, such as *VQE* [39] or *QAOA* [23], perform several iterations of quantum and classical processing until the result converges [56]. Thus, the quantum and classical programs have to be integrated to retrieve the final result.

The different programs of QAIs, as well as the QAIs and classical programs of hybrid quantum applications, have to be orchestrated and required data must be passed between them [85, 90]. Workflow technology is an orchestration approach that has been proven since decades to be applicable in various heterogeneous application areas [50, 53]. Hence, workflows should also be used for orchestrating the programs constituting a quantum application [47]. For this, the required *activities* invoking the quantum and classical programs, their execution order, and the data flow between them, are specified in so-called *workflow models* [20, 51]. Such workflow models can automatically be executed by a *workflow engine*. In contrast to the orchestration using a traditional program, e.g., written in Java or Python, workflows provide different benefits, such as robustness, scalability, or persistence [51, 90]. Further, alternative control flows in the presence of errors, as well as transactions comprising multiple activities, can be defined [37]. Thus, hybrid quantum applications will benefit from the usage of one or multiple workflow models orchestrating the required programs.

## 3 Quantum Software Development Lifecycle

In this section, we present our quantum software development lifecycle, which is depicted in Figure 2. For this, we first discuss that the development of hybrid quantum applications requires integrating the lifecycles of different software artifacts. Then, we present the various phases of the quantum software development lifecycle.

### 3.1 Interwoven Lifecycles

As discussed in the previous section, quantum applications are usually compound from different artifacts, namely quantum and classical programs and workflows to orchestrate them [47, 90]. Thus, in addition to phases, such as the requirement analysis or the design of the application, the development of a quantum application also comprises the development of the constituting software artifacts. Hence, the quantum software development lifecycle incorporates multiple lifecycles that are interwoven, as depicted in Figure 2: (i) the *quantum workflow lifecycle*, (ii) the *classical software lifecycle*, and (iii) the *quantum circuit lifecycle*. Further, the various software artifacts also have to be managed, which is prepared and done in the (iv) *operations lifecycle*. Thereby, developers and operations personnel should be tightly integrated following the widely used *DevOps paradigm* [7, 29] to enable fast and frequent releases [92]. This is especially important in the quantum computing domain with the rapid development of new quantum computers or software tools, which may require adapting the quantum applications regularly [81]. Hence, these lifecycles have to be integrated into the overall lifecycle, and concepts, best practices, and tools used in the various lifecycles must be considered when developing hybrid quantum applications.

There are various lifecycles for business process management and workflows proposed by different works [19, 42, 51]. We base our lifecycle for quantum workflows on the lifecycles presented by Leymann and Roller [51], as well as Dumas et al. [19]. Furthermore, we added required phases specific to the quantum computing domain. The quantum workflow lifecycle is discussed in detail in Section 3.3.

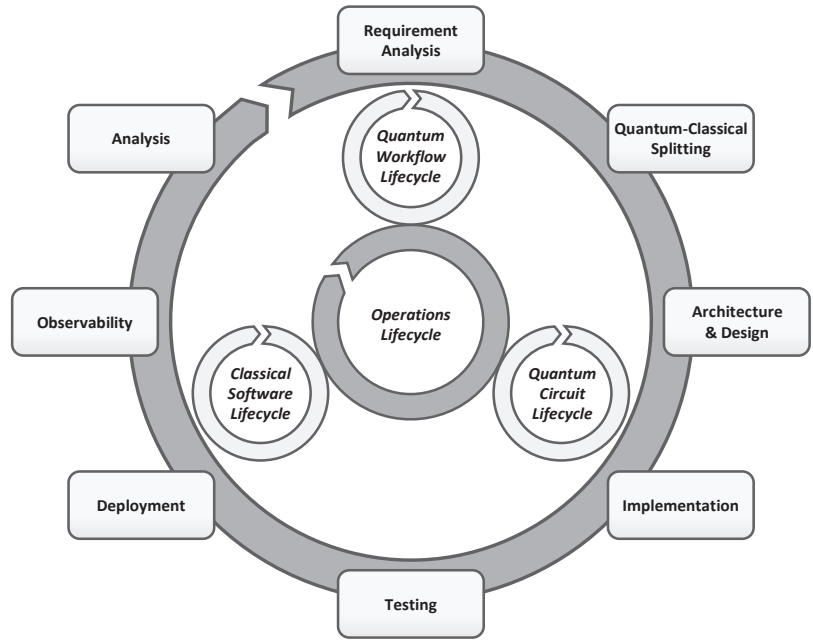

**Fig. 2** Overview of the Quantum Software Development Lifecycle.

Similar to workflow lifecycles also multiple lifecycles for the development of classical software artifacts have been introduced [14, 55, 59]. A widely used software lifecycle is the *waterfall model*, comprising five phases: (i) requirement analysis, (ii) design, (iii) implementation, (iv) testing, and (v) maintenance [59]. Other lifecycles or software development models are the *spiral model* [10], the *V-model* [55], and the *prototype model* [43]. However, the detailed discussion of lifecycles for the development of classical software artifacts is out of the scope of this chapter.

Additionally, a lifecycle for the development of quantum circuits has to be integrated into the overall lifecycle for the development of quantum applications. Thereby, we use the quantum circuit lifecycle that we proposed in previous work [86], which will be discussed in more detail in Section 3.4. Although there are some other lifecycles [18, 97], they are still abstract and need to be refined to capture all relevant details to guide developers (see Section 5). However, also lifecycles documenting the relevant phases in the development of quantum programs for other quantum computing models, e.g., the adiabatic model [2], can be integrated in the future.

Finally, all the developed software artifacts constituting a quantum application have to be operated. This includes, e.g., the packaging of the quantum application to ship it into the target environment, or its deployment [7]. The required concepts and tools differ from classical DevOps and have to be extended for the quantum computing domain [29]. Section 3.5 presents the operations lifecycle exhaustively.

## 3.2 Enclosing Lifecycle

In the following, we introduce the enclosing lifecycle depicted in Figure 2, defining how the different lifecycles must be interwoven. Thus, various phases require entering the lifecycles of the software artifacts constituting the quantum application, e.g., the implementation phase. Other phases rely on the interplay of the corresponding phases from the different lifecycles, e.g., the testing and deployment phases. Hence, the goal of these phases is summarized, and the phases that must be integrated are discussed.

### 3.2.1 Requirement Analysis

For both classical and quantum application development, the different interested stakeholders must identify their *requirements* first [18, 32]. Thereby, the requirements can be functional, i.e., defining the problems to solve, and non-functional, i.e., specifying quality attributes of the resulting quantum application, such as availability, scalability, performance, or maintainability [76]. The requirements are documented measurably to enable the evaluation of the resulting quantum application in the later lifecycle phases, e.g., the analysis phase (see Section 3.2.8) [97]. Further, the different requirements are prioritized, and the overall project schedule is elaborated [7, 32].

### 3.2.2 Quantum-Classical Splitting

The *quantum-classical splitting* phase is intended to decide which parts of the quantum application to execute on a quantum computer and which on a classical computer [64, 86]. For the quantum parts, it is also determined if, e.g., a gate-based quantum computer [44] or a quantum annealer [2] should be used. The splitting is based on the requirements from the previous phase, i.e., it is evaluated for which parts suited quantum algorithms exist [18]. Furthermore, it is verified if the non-functional requirements can be satisfied by a quantum program considering the capabilities of the available quantum computers [71]. The splitting can, e.g., be done by quantum experts based on their knowledge and experience [86]. However, this task is complex, time-consuming, and error-prone. Therefore, it should be automated or supported by a recommendation system, which can be based on *patterns* [45, 91] and best practices or so-called *provenance data* [33, 87] about other quantum applications.

### 3.2.3 Architecture & Design

The result of the previous phase is a collection of quantum and classical parts. In the *architecture & design* phase, an architecture is conceptualized by using these parts and specifying corresponding software components with their functionality and interfaces [7, 59]. Then, the architecture is refined with the internals of the different software components, e.g., the used data structures [97]. The resulting description should provide enough details for the implementation of the various components in the next phase. Thereby, it can, e.g., be specified using the *Unified Modeling Language (UML)*, for which an extension for quantum computing exists [28, 65].

### 3.2.4 Implementation

In the next phase, the quantum application is implemented based on the requirements and design from the previous phases. Thereby, the implementation includes the development of the different constituting software artifacts. This means the lifecycles for classical programs and quantum programs (see Section 3.4) are entered in this phase. Furthermore, workflows should be used for orchestrating the control- and data flow between these programs [47, 90]. Thus, also the quantum workflow lifecycle is interwoven into this phase (see Section 3.3). The reuse of existing code and programs is one of the major goals of quantum software engineering [66, 67]. Hence, before entering the lifecycles to develop the required software artifacts, existing code and implementations are searched, e.g., using an API manager [17], a service registry [27], or a platform for sharing quantum software [48, 49].

### 3.2.5 Testing

After the implementation, the quantum application is tested to verify the intended behavior according to the specified requirements before delivering it to the users [59, 97]. Similar to the implementation phase, it includes the testing of all constituting software artifacts, i.e., quantum programs, classical programs, and workflows. Therefore, the testing of these artifacts is also located in their corresponding lifecycles (e.g., see Section 3.4.3). In addition to testing the artifacts in isolation, so-called *integration tests* should be performed to verify if the independently developed artifacts work together correctly [94]. Although there are some testing and verification approaches for quantum circuits, the development of a holistic testing strategy for hybrid quantum applications is still an open research question [4, 58, 83].

### 3.2.6 Deployment

During the *deployment* phase, everything is prepared to enable the execution of the quantum application [93, 95]. Thus, the execution environment for the classical programs is set up, e.g., for a Python script, a virtual machine may be created and, the required Python runtime is installed on it [88]. Similarly, also the quantum programs and the workflows must be deployed. However, some of the required functionality may also be available as a service or API and require no deployment [89]. The deployment is part of the operations lifecycle and is discussed in detail in Section 3.5.4.

### 3.2.7 Observability

In the next phase, the quantum application, as well as its execution environment, are monitored. Thereby, data is collected for two different purposes: (i) observing the current state of a running quantum application and (ii) storing the data in the long-term to enable its analysis, e.g., to improve the quantum application or to enable traceability, comprehensibility, and reproducibility [33, 51]. This phase requires the collection of data about all software artifacts comprising the hybrid quantum application [87]. Hence, it must be defined in the different development lifecycles what data to collect, which is then gathered at runtime in the operations lifecycle.

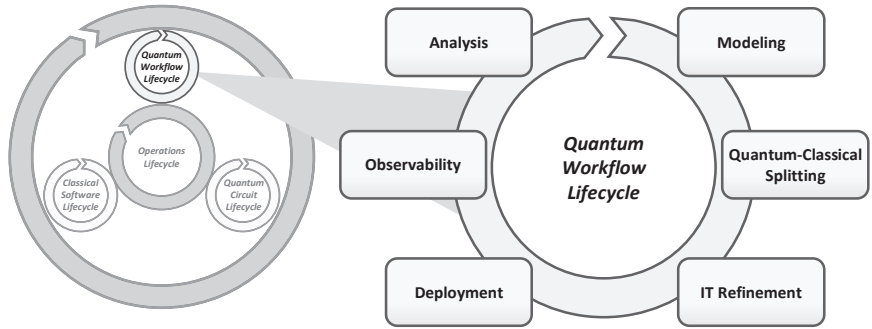

**Fig. 3** Detailed View of the Quantum Workflow Lifecycle.

### 3.2.8 Analysis

In the last phase of the lifecycle, the collected data from the observability phase is analyzed. The goals of this phase are to find bugs that have to be fixed or possible improvements for the quantum application [47, 87]. For example, if the quantum programs frequently produce erroneous results, a sub-optimal splitting for today's limited quantum computers could be the reason [69, 86]. Therefore, after the analysis phase, the next iteration of the lifecycle can be entered, e.g., adapting the requirements to perform an improved splitting and realize the other found optimizations.

## 3.3 Quantum Workflow Lifecycle

As discussed in Section 2, the different programs realizing a quantum application have to be orchestrated, which should be done using workflows to benefit from their advantages [47, 88]. Next, we present the quantum workflow lifecycle (see Figure 3).

### 3.3.1 Modeling

In the *modeling* phase, the collection of activities implementing the quantum application, as well as their partial order, are defined in a workflow model depending on the result of the architecture & design phase (see Section 3.2.3) [47, 51]. Furthermore, also the data flow between the activities is specified [20, 37]. Thereby, a standardized workflow language, such as the *Business Process Model and Notation (BPMN)* [63] or the *Business Process Execution Language (BPEL)* [61], should be used to simplify the reuse of workflow models across different workflow engines [50, 51].

### 3.3.2 Quantum-Classical Splitting

Similar to the quantum-classical splitting of the enclosing lifecycle (see Section 3.2.2), a splitting is also performed in the quantum workflow lifecycle. Thereby, the goal is to decide which of the activities of the workflow model from the previous phase are implemented classically and which require the execution of a quantum algorithm or program. For this, corresponding extensions for workflow languages have been proposed providing explicit modeling constructs for the execution of quantum circuits, as well as frequently occurring pre- and post-processing tasks [85, 90].

### 3.3.3 IT Refinement

The *IT refinement* phase is intended to transform the abstract workflow model from the previous phases into an executable workflow model [51]. For this, the contained activities are refined regarding three dimensions: (i) *what* has to be done within the activity, (ii) *with* which programs is the activity performed, and (iii) *who* is responsible for the activity. Thereby, existing implementations for the activities should be searched first, e.g., quantum and classical programs or workflow models that can be used as sub-workflows to increase the software reuse [17, 27, 66]. If no suited implementation to reuse is found, it must be implemented in this phase by entering the corresponding lifecycle, e.g., the quantum circuit lifecycle (see Section 3.4).

### 3.3.4 Deployment

In the *deployment* phase, the modeled and refined workflow model is uploaded to the workflow engine [19, 51]. Thereby, the workflow model is usually frozen, i.e., it can no longer be changed. Thus, the upload of a changed workflow model from another iteration of the lifecycle results in a new version of the workflow model and does not affect running instances. The implementations of the different activities in the workflow can either be bound during deployment or dynamically at runtime [27]. After the deployment, the workflow is ready for execution and can be instantiated.

### 3.3.5 Observability

The created workflow instances are monitored to track their current state during runtime [20, 51]. This includes, e.g., the currently executed activities, the input and output data of already performed activities, or the reason for taking a particular path in the workflow model [47, 50]. The collected information can usually be visualized by the workflow engine and, e.g., used to handle unexpected errors [51]. When a workflow instance terminates, the collected data is moved to the *audit trail*, a separate database comprising the information about completed workflow instances [1, 84].

### 3.3.6 Analysis

In the last phase, the data stored in the audit trail is analyzed, e.g., using process mining or machine learning techniques [1, 68]. Thereby, statistics about the various paths taken through the workflow model or the average execution times can be used as a basis for redesigning and improving the workflow model in the next iteration [51]. This redesign, e.g., includes parallelizing activities, adding automated error handling for frequently occurring errors, or improving slow activity implementations.

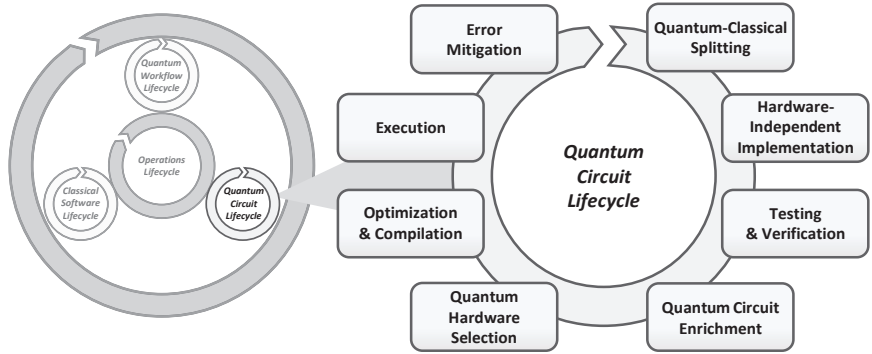

**Fig. 4** Detailed View of the Quantum Circuit Lifecycle.

### 3.4 Quantum Circuit Lifecycle

In the following, we discuss our quantum circuit lifecycle [86], as depicted in Figure 4. This lifecycle is, e.g., entered if a quantum circuit is required as part of a quantum application, and no suitable implementation can be found (see Section 3.2.4).

#### 3.4.1 Quantum-Classical Splitting

The splitting in the quantum circuit lifecycle is the splitting at the lowest granularity compared to the splitting of quantum applications and quantum workflows. It is entered with a description of the problem to solve and is intended to decide if a pure quantum algorithm or a hybrid quantum algorithm should be used [86]. For example, if the problem is to find eigenvalues, the *quantum phase estimation (QPE)* as a pure quantum algorithm or the *variational quantum eigensolver (VQE)* as a hybrid algorithm can be used [82], and the decision is done in this lifecycle phase.

#### 3.4.2 Hardware-Independent Implementation

After deciding which quantum algorithm to use, the corresponding quantum circuit must be implemented. Thereby, the implementation should be hardware-independent to enable a later hardware selection based on the current characteristics of the different available quantum computers (see Section 3.4.5) [71, 79, 85]. Furthermore, the quantum circuit should also be defined independently of specific input data, which is encoded in the next phase by prepending a suited state preparation circuit to the beginning of the implemented circuit [16, 91]. Thus, the quantum circuit can be reused for different instances of the problem to solve [86]. For the implementation of the quantum circuit, a plethora of technologies can be utilized [24, 46, 81]. For example, (i) quantum programming languages, such as *Q#* or *Quipper*, (ii) quantum assembly languages, such as *OpenQASM* or *Quil*, and (iii) quantum libraries that are embedded into classical programming languages, such as *Qiskit* or *Forest* in Python.

### 3.4.3 Testing & Verification

Next, the quantum circuit is tested and verified to ensure its correct behavior. One approach is to add statistical assertions to the quantum circuit [35]. Then, it is verified that the specified state is measured when executing the quantum circuit until the point where the assertion is defined. Thus, the results of the assertions guide programmers in finding bugs. However, this requires the execution of the circuit for each assertion, which is only feasible for small quantum circuits and few assertions. Additionally, first approaches try to check assertions dynamically at runtime [52]. But then additional ancilla qubits and gates are required, limiting the applicability with today's restricted quantum computers [69]. Another approach adapted from classical software engineering is white- and black-box testing, e.g., utilizing a simulator if this is feasible for the quantum circuit size [58, 80]. Further, quantum circuits can also be verified by experts or using automated approaches [4, 83]. However, the debugging, testing, and verification of quantum circuits is still an open research question.

### 3.4.4 Quantum Circuit Enrichment

The quantum circuit from the hardware-independent implementation phase is implemented independently of a certain problem instance to solve. Thus, it is enriched with the details required to solve a particular instance of the problem in this phase [86]. This enrichment comprises two steps: (i) *state preparation* [16, 91] and (ii) *oracle expansion* [40]. For the state preparation step, a circuit initializing the register of the quantum computer with the required state is generated based on the input data [46, 91]. The resulting state preparation circuit is then prepended to the original circuit. Thereby, different encodings exist, such as the *angle*, *amplitude*, or *basis encoding* [91]. These encodings provide different characteristics, e.g., the number of required qubits or gates. Furthermore, different quantum algorithms rely on black-box functions, so-called *oracles* [40, 46]. However, these oracles have to be implemented or loaded from a corresponding library before executing the quantum circuit.

### 3.4.5 Quantum Hardware Selection

Quantum computers that are available during the *Noisy Intermediate-Scale Quantum (NISQ)* [69] era are error-prone and provide only limited capabilities [46, 71]. Additionally, periodic re-calibrations change their characteristics, e.g., the decoherence times of the qubits, over time [79, 87]. Thus, the selection of a suitable quantum computer to execute a given quantum circuit is a complex task [71, 85]. To overcome this issue, different metrics, such as *quantum volume (QV)* [9] or the *total quantum factor (TQF)* [72], and various benchmarks [41, 57] have been introduced to assess the capabilities of the available quantum computers. Further, there are some tools, such as the *QuRE Toolbox* [78], to estimate the required resources to execute a quantum algorithm on given input data. Finally, the *NISQ Analyzer* [71] automatically selects a suitable quantum computer based on properties of the quantum circuit, such as width or depth, and the current characteristics of the available quantum computers.

### 3.4.6 Optimization & Compilation

After selecting a suitable quantum computer for the execution of the quantum circuit, it has to be optimized and compiled to the machine instructions that can be executed by the selected quantum computer [11, 75]. For this, a quantum compiler assigns the qubits of the quantum circuit to the physical qubits of the quantum computer [34, 38, 46]. Due to the different characteristics of the qubits, e.g., their decoherence times or connectivity, the assignments influence the error probability of the quantum circuit execution [46]. Therefore, the assignments should be optimized based on current provenance data about the qubit characteristics [87]. Similarly, the gates used in the quantum circuit must be mapped to gates physically implemented by the selected quantum computer [11]. If one of the gates is not physically implemented, it has to be replaced by a corresponding subroutine of implemented gates [75, 87].

### 3.4.7 Execution

In the next phase, the compiled quantum circuit is executed on the selected quantum computer. Depending on the quantum cloud offering used, this is done by submitting a corresponding job to a queue or reserving a time slice for the execution [44, 49]. The quantum circuit is usually executed multiple times, referred to as the number of shots, to reduce the impact of statistical errors [46, 87]. Furthermore, if a variational algorithm is selected in the quantum-classical splitting phase, the execution may comprise multiple iterations of quantum and classical processing [46, 86].

### 3.4.8 Error Mitigation

In contrast to full *error correction* [26, 70], which is unfeasible on today's NISQ machines, *error mitigation* [77] has the goal to reduce the impact of noise in the results of quantum circuit executions [46]. Some of these error mitigation techniques require adding additional gates or adapting existing ones while using much fewer qubits than needed for error correction [21]. However, the circuit then has to be adapted before the execution. After the execution, classical post-processing is used to mitigate the errors [77]. Some techniques also solely rely on classical post-processing and do not require changes in the circuits [22]. A subset of these techniques are so-called readout-error mitigation or unfolding techniques [13, 54]. Thereby, depending on the used technique, different states are periodically prepared and subsequently measured on the quantum computer [86]. Based on the retrieved data, the impact of readout-errors can then be reduced in the result distribution of a circuit execution [13].

## 3.5 Operations Lifecycle

The last lifecycle integrated into the quantum software development lifecycle is the operations lifecycle, for which the different phases are depicted in Figure 5. It is intended to operate all the software artifacts comprising a quantum application.

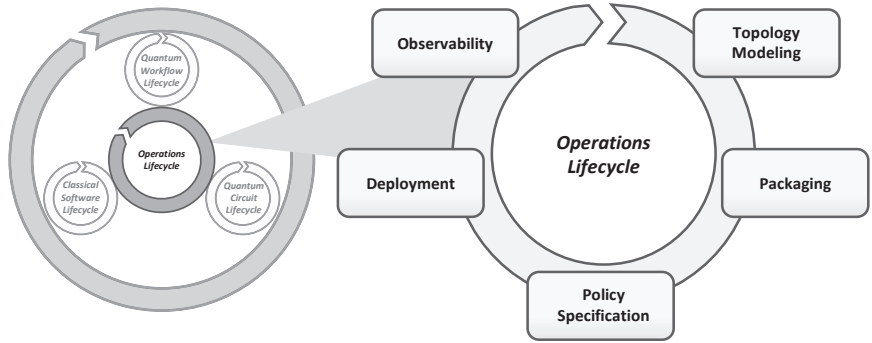

**Fig. 5** Detailed View of the Operations Lifecycle.

### 3.5.1 Topology Modeling

The operations personnel performing the phases in this lifecycle are in charge of deploying and managing all software artifacts of the hybrid quantum application (see Section 3.5.4). However, a manual deployment and management are time-consuming and error-prone [12, 89]. Thus, it must be automated using so-called *provisioning* or *deployment technologies*, such as Kubernetes or Terraform [93, 95]. For this, all necessary software artifacts and their dependencies are described by a directed acyclic graph, called the *topology model* [8]. In addition to the proprietary languages provided by the different provisioning technologies, there are also standardized languages such as *TOSCA* [62] to define topology models [93]. Figure 6 depicts an exemplary topology model for a hybrid quantum application. Thereby, the nodes in the topology model represent the different software artifacts, e.g., the classical and quantum programs [88]. Further, the edges specify the relations between the artifacts, e.g., that the classical program is hosted on a docker engine or connects to a quantum program after performing some pre-processing. The semantics of the nodes is defined by reusable types shown in brackets [95], e.g., the quantum program is implemented as a Qiskit app and executed using the IBMQ quantum cloud offering. Finally, the nodes in the topology model can be configured using so-called properties, e.g., the token to access IBMQ at runtime as shown at the corresponding node [93].

### 3.5.2 Packaging

After specifying the topology model, the quantum application is packaged as a self-contained archive [29, 47]. Therefore, only a single entity including all dependencies has to be transferred into the target environment for the execution [89]. This self-contained archive contains the quantum and classical programs comprising the quantum application, as well as the topology model from the previous phase describing their dependencies and how they can be automatically provisioned [47, 89]. Furthermore, workflow models to orchestrate the programs can be added to the archive. Finally, data required by the quantum application may also be packaged [98].

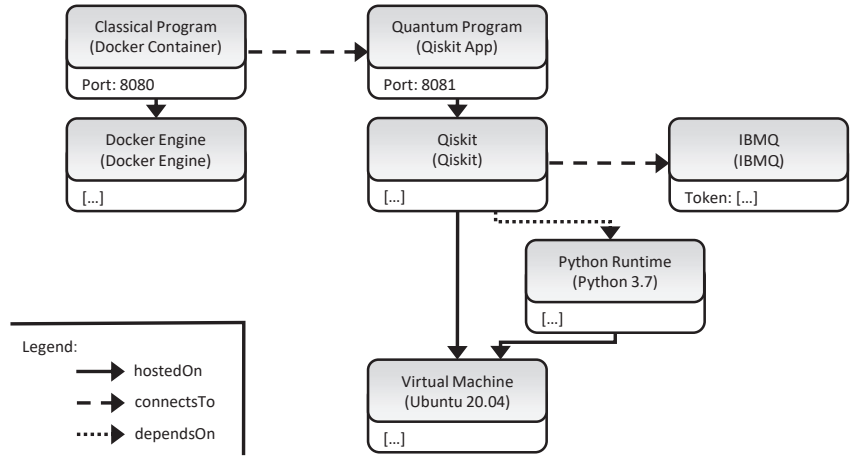

**Fig. 6** Exemplary Topology Model for a Hybrid Quantum Application (based on [93]).

### 3.5.3 Policy Specification

The developed quantum application can usually be offered with different quality of service (QoS) guarantees [15]. For example, the classical components of the quantum application can be automatically scaled, or a defined time slice can be reserved for the quantum programs. Therefore, different policies can be defined specifying the QoS guarantees as well as the implications when using the policy, e.g., the incurred monetary costs, to offer them in an app store or over an API manager [17, 49].

### 3.5.4 Deployment

In this phase, the execution environment for the quantum application is set up. For this, the topology model is passed to a corresponding provisioning engine, which interprets it and installs the required dependencies and programs [47, 95]. In addition to a deployment for all users if a new version is available, also advanced strategies, such as performing a *canary deployment* [3], are possible. This allows deploying the new version for a subset of users to evaluate it before rolling it out for all users.

### 3.5.5 Observability

During runtime, the deployed software artifacts are monitored to verify their correct behavior or to visualize their current state for the user (see Section 3.3.5). Thereby, the collected data for all software artifacts constituting the quantum application must be consolidated to enable a unified view [87]. This comprises, e.g., the logs of a virtual machine executing a classical program, the logs of a workflow instance, or the current characteristics of the used quantum computers. Furthermore, this data is stored in the long-term to enable the analysis of the quantum application (see Section 3.2.8).

## 4 Discussion

The introduced quantum software development lifecycle integrates the quantum workflow lifecycle, implying that most non-trivial quantum applications should be implemented using workflow technology [47, 90]. Thereby, workflow technology enables benefiting from robust, proven, and mature solutions, that have been applied in various heterogeneous application areas, such as *e-Science* [53] or *business process management* [50]. Furthermore, there are already the first commercial workflow offerings specialized for quantum computing like Zapata *Orquestra* [96]. Finally, IBM announced workflows as one of the major building blocks in their roadmap [36]. However, our lifecycle can also be used without workflows if it turns out in the architecture & design phase that workflows are not required for the application.

The presented quantum circuit lifecycle (see Section 3.4) is designated for the development and execution of quantum circuits during the NISQ era [69, 86]. Therefore, it contains some phases that can be skipped if fully error-corrected quantum computers are available, e.g., the hardware selection or error mitigation phase. Furthermore, other lifecycle phases might change significantly due to new developments. For example, the quantum circuit enrichment phase must be adapted if an efficient *quantum random access memory (QRAM)* [31] implementation is available [86, 91].

Additionally, the quantum software development lifecycle is assuming gate-based quantum algorithm implementations. However, it can be easily adapted by integrating a lifecycle for quantum programs relying on other quantum computing models, e.g., the adiabatic model [2], as discussed in Section 3.1. To the best of our knowledge, there exists currently no lifecycle for quantum programs using another model.

Finally, there are a lot of open research questions and possibilities to improve the lifecycle phases. For example, developing a holistic test strategy for hybrid quantum applications, the proposal of new metrics to assess and compare quantum computers, or designing a recommendation system for the quantum-classical splitting phases.

## 5 Related Work

Different research works proposed lifecycles, methodologies, or workflows for the development of quantum applications which will be discussed in this section.

Zhao [97] performed a comprehensive survey about quantum software engineering, presenting different methods, tools, and open questions in this research area. Additionally, he also introduces a quantum software lifecycle based on the classical waterfall model, consisting of five phases: (i) *quantum software requirements analysis*, (ii) *quantum software design*, (iii) *quantum software implementation*, (iv) *quantum software testing*, and (v) *quantum software maintenance*. Thereby, the different phases are reused from the classical lifecycle, but the tools and methods for the phases are adapted to the quantum computing domain. However, it misses the discussion of some important aspects, such as the deployment of quantum applications, the orchestration of the quantum and classical programs, and the packaging of all required artifacts, e.g., to store and sell the quantum application in an app store.

A quantum software lifecycle similar to Zhao's is also proposed by Dey et al. [18]. Thereby, they include the same five phases but add an additional *quantum feasibility study* phase to the beginning of the lifecycle. This phase is intended to evaluate the availability of suited quantum algorithms, as well as powerful enough quantum computers. In our lifecycle, this is included in the quantum-classical splitting phase of the enclosing lifecycle, which separates the problem into classical parts and quantum parts that can be successfully executed on an available quantum computer. However, it also does not include some important phases, e.g., the monitoring of the running quantum application or their packaging as a self-contained archive and deployment.

Quantum DevOps was proposed by Gheorghe-Pop et al. [29], motivating the need to apply the DevOps paradigm in the quantum computing domain. Thereby, they analyze the different phases of the traditional DevOps process and extend them correspondingly. Further, they focus on the evaluation of the available quantum computers in each iteration, to enable the selection of a suitable one for the execution.

Sodhi et al. [76] analyzed the characteristics of different quantum computing platforms, e.g., from IBM and Rigetti. Based on this analysis, they examined how the characteristics affect quality attributes of quantum applications, such as maintainability, usability, or performance. Further, the impact on the various lifecycle phases and required steps to achieve the quality attributes in these phases are discussed.

## 6 Conclusion and Outlook

Quantum computers are rapidly evolving in terms of qubit counts, decoherence times, and lower error rates. Thus, problems in more and more application areas can be solved by quantum applications. Hence, the need for high-quality quantum applications will increase dramatically in the next years. However, the development of such applications is complex and incorporates experts from various fields. To enable their successful cooperation and ease the education of new developers, a common understanding of the development process of quantum applications is needed. In this chapter, we introduced a quantum software development lifecycle summarizing eight phases comprising this development process. Furthermore, we discussed the different software artifacts usually realizing a quantum application, i.e., quantum programs, classical programs, and workflows. We presented the lifecycles of these artifacts and showed how they are integrated into the overall lifecycle of quantum applications.

Quantum computing in general, and also quantum software engineering, is a very active research area where new concepts and tools are published regularly. Therefore, the quantum software development lifecycle is a living document, which can be adapted and extended with new developments. This comprises, e.g., the addition of new concepts and tools or the extension with another development phase.

**Acknowledgements** This work was funded by the BMWi project *PlanQK* (01MK20005N), the DFG's Excellence Initiative project *SimTech* (EXC 2075 – 390740016), and the project *SEQUOIA* funded by the Baden-Wuerttemberg Ministry of the Economy, Labour and Housing.

Quantum Software Development Lifecycle 17## References

1. Agrawal, R., Gunopulos, D., Leymann, F.: Mining Process Models from Workflow Logs. In: International Conference on Extending Database Technology, pp. 467–483. Springer (1998)
2. Aharonov, D., Van Dam, W., Kempe, J., Landau, Z., Lloyd, S., Regev, O.: Adiabatic Quantum Computation Is Equivalent to Standard Quantum Computation. SIAM review **50**(4), 755–787 (2008)
3. Ahmadighohandizi, F., Systä, K.: Application Development and Deployment for IoT Devices. In: Proceedings of the 4$^{th}$ European Conference on Service-Oriented and Cloud Computing (ESOCC), pp. 74–85. Springer (2016)
4. Amy, M.: Towards Large-scale Functional Verification of Universal Quantum Circuits. arXiv:1805.06908 (2018)
5. Barzen, J.: From Digital Humanities to Quantum Humanities: Potentials and Applications. In: Quantum Computing in the Arts and Humanities. Springer (2021). arXiv:2103.11825
6. Barzen, J., Leymann, F., Falkenthal, M., Vietz, D., Weder, B., Wild, K.: Relevance of Near-Term Quantum Computing in the Cloud: A Humanities Perspective. Cloud Computing and Services Science **1399**, 25–58 (2021)
7. Bass, L., Weber, I., Zhu, L.: DevOps: A Software Architect's Perspective. Addison-Wesley Professional (2015)
8. Binz, T., Breiter, G., Leymann, F., Spatzier, T.: Portable Cloud Services Using TOSCA. IEEE Internet Computing **16**(3), 80–85 (2012)
9. Bishop, L.S., et al.: Quantum volume. Technical Report (2017)
10. Boehm, B.W.: A spiral model of software development and enhancement. Computer **21**(5), 61–72 (1988)
11. Booth Jr, J.: Quantum Compiler Optimizations. arXiv:1206.3348 (2012)
12. Breitenbücher, U., Binz, T., Képes, K., Kopp, O., Leymann, F., Wettinger, J.: Combining Declarative and Imperative Cloud Application Provisioning based on TOSCA. In: International Conference on Cloud Engineering (IC2E), pp. 87–96. IEEE (2014)
13. Brenner, L., Verschuuren, P., Balasubramanian, R., Burgard, C., Croft, V., Cowan, G., Verkerke, W.: Comparison of unfolding methods using RooFitUnfold. arXiv:1910.14654 (2019)
14. Canós, J.H., Penadés, M.C., Carsí, J.Á.: From Software Process to Workflow Process: the Workflow Lifecycle. In: Proceedings of the International Process Technology Workshop (1999)
15. Cardoso, J., Sheth, A., Miller, J., Arnold, J., Kochut, K.: Quality of service for workflows and web service processes. Journal of web semantics **1**(3), 281–308 (2004)
16. Cortese, J.A., Braje, T.M.: Loading Classical Data into a Quantum Computer. arXiv:1807.02500 (2018)
17. De, B.: Api Management. In: API Management, pp. 15–28. Springer (2017)
18. Dey, N., Ghosh, M., kundu, S.S., Chakrabarti, A.: QDLC–The Quantum Development Life Cycle. arXiv:2010.08053 (2020)
19. Dumas, M., La Rosa, M., Mendling, J., Reijers, H.A.: Fundamentals of Business Process Management, vol. 1. Springer (2013)
20. Ellis, C.A.: Workflow Technology. Computer Supported Cooperative Work, Trends in Software Series **7**, 29–54 (1999)
21. Endo, S., Benjamin, S.C., Li, Y.: Practical Quantum Error Mitigation for Near-Future Applications. Physical Review X **8**(3), 031027 (2018)
22. Endo, S., Cai, Z., Benjamin, S.C., Yuan, X.: Hybrid Quantum-Classical Algorithms and Quantum Error Mitigation. Journal of the Physical Society of Japan **90**(3), 032001 (2021)
23. Farhi, E., Goldstone, J., Gutmann, S.: A Quantum Approximate Optimization Algorithm. arXiv:1411.4028 (2014)
24. Fingerhuth, M., Babej, T., Wittek, P.: Open source software in quantum computing. PloS one **13**(12) (2018)
25. Gabor, T., et al.: The Holy Grail of Quantum Artificial Intelligence: Major Challenges in Accelerating the Machine Learning Pipeline. arXiv:2004.14035 (2020)

26. Gaitan, F.: Quantum error correction and fault tolerant quantum computing. CRC Press (2008)
27. Garofalakis, J., Panagis, Y., Sakkopoulos, E., Tsakalidis, A.: Contemporary Web Service Discovery Mechanisms. Journal of Web Engineering **5**(3), 265–290 (2006)
28. Gemeinhardt, F., Garmendia, A., Wimmer, M.: Towards Model-Driven Quantum Software Engineering. In: Proceedings of the 2nd International Workshop on Quantum Software Engineering (Q-SE). ACM (2021)
29. Gheorghe-Pop, I.D., Tcholtchev, N., Ritter, T., Hauswirth, M.: Quantum DevOps: Towards Reliable and Applicable NISQ Quantum Computing. In: IEEE Globecom Workshops, pp. 1–6. IEEE (2020)
30. Ghezzi, C., Jazayeri, M., Mandrioli, D.: Fundamentals of Software Engineering (2002)
31. Giovannetti, V., Lloyd, S., Maccone, L.: Quantum random access memory. Physical review letters **100**(16), 160501 (2008)
32. Grady, J.O.: System Requirements Analysis. Elsevier (2010)
33. Herschel, M., Diestelkämper, R., Ben Lahmar, H.: A Survey on Provenance: What for? What Form? What from? The VLDB Journal **26**(6), 881–906 (2017)
34. Heyfron, L.E., Campbell, E.T.: An efficient quantum compiler that reduces T count. Quantum Science and Technology **4**(1), 015004 (2018)
35. Huang, Y., Martonosi, M.: Statistical Assertions for Validating Patterns and Finding Bugs in Quantum Programs. In: Proceedings of the 46th International Symposium on Computer Architecture, pp. 541–553. ACM (2019)
36. IBM: IBM's roadmap for building an open quantum software ecosystem (2021). URL https://www.ibm.com/blogs/research/2021/02/quantum-development-roadmap
37. J. Eder and W. Liebhart: Workflow transactions. Workflow Handbook pp. 195–202 (1997)
38. JavadiAbhari, A., et al.: ScaffCC: A Framework for Compilation and Analysis of Quantum Computing Programs. In: Proceedings of the 11th Conference on Computing Frontiers, pp. 1–10. ACM (2014)
39. Kandala, A., et al.: Hardware-efficient variational quantum eigensolver for small molecules and quantum magnets. Nature **549**(7671), 242–246 (2017)
40. Kashefi, E., Kent, A., Vedral, V., Banaszek, K.: Comparison of quantum oracles. Physical Review A **65**(5), 050304 (2002)
41. Knill, E., Laflamme, R., Martinez, R., Negrevergne, C.: Benchmarking Quantum Computers: The Five-Qubit Error Correcting Code. Physical Review Letters **86**, 5811–5814 (2001)
42. Kohlborn, T., Korthaus, A., Rosemann, M.: Business and Software Service Lifecycle Management. In: Proceedings of the 13th International Enterprise Distributed Object Computing Conference (EDOC), pp. 87–96. IEEE (2009)
43. Kumar, N., Zadgaonkar, A., Shukla, A.: Evolving a New Software Development Life Cycle Model SDLC-2013 with Client Satisfaction. International Journal of Soft Computing and Engineering (IJSCE) **3**(1), 2231–2307 (2013)
44. LaRose, R.: Overview and Comparison of Gate Level Quantum Software Platforms. Quantum **3**, 130 (2019)
45. Leymann, F.: Towards a pattern language for quantum algorithms. In: Quantum Technology and Optimization Problems, pp. 218–230. Springer International Publishing (2019)
46. Leymann, F., Barzen, J.: The bitter truth about gate-based quantum algorithms in the nisq era. Quantum Science and Technology **5**(4), 044007 (2020)
47. Leymann, F., Barzen, J.: Hybrid Quantum Applications Need Two Orchestrations in Superposition: A Software Architecture Perspective. arXiv:2103.04320 (2021)
48. Leymann, F., Barzen, J., Falkenthal, M.: Towards a Platform for Sharing Quantum Software. In: Proceedings of the 13th Advanced Summer School on Service-Oriented Computing (SummerSOC), IBM Technical Report, pp. 70–74. IBM Research Division (2019)
49. Leymann, F., Barzen, J., Falkenthal, M., Vietz, D., Weder, B., Wild, K.: Quantum in the Cloud: Application Potentials and Research Opportunities. In: Proceedings of the 10th International Conference on Cloud Computing and Services Science (CLOSER), pp. 9–24. SciTePress (2020)
50. Leymann, F., Roller, D.: Workflow-based applications. IBM Systems Journal **36**(1), 102–123 (1997)

51. Leymann, F., Roller, D.: Production Workflow: Concepts and Techniques. Prentice Hall PTR (2000)
52. Liu, J., Byrd, G.T., Zhou, H.: Quantum Circuits for Dynamic Runtime Assertions in Quantum Computation. In: Proceedings of the 25$^{th}$ International Conference on Architectural Support for Programming Languages and Operating Systems, pp. 1017–1030. ACM (2020)
53. Liu, J., Pacitti, E., Valduriez, P., Mattoso, M.: A Survey of Data-Intensive Scientific Workflow Management. Journal of Grid Computing **13**(4), 457–493 (2015)
54. Maciejewski, F.B., et al.: Mitigation of readout noise in near-term quantum devices by classical post-processing based on detector tomography. Quantum **4** (2020)
55. Mathur, S., Malik, S.: Advancements in the V-Model. International Journal of Computer Applications **1**(12), 29–34 (2010)
56. McClean, J.R., Romero, J., Babbush, R., Aspuru-Guzik, A.: The theory of variational hybrid quantum-classical algorithms. New Journal of Physics **18**(2), 023023 (2016)
57. Michielsen, K., Nocon, M., Willsch, D., Jin, F., Lippert, T., De Raedt, H.: Benchmarking gate-based quantum computers. Computer Physics Communications **220**, 44 – 55 (2017)
58. Miranskyy, A., Zhang, L., Doliskani, J.: Is Your Quantum Program Bug-Free? In: Proceedings of the ACM/IEEE 42$^{nd}$ International Conference on Software Engineering: New Ideas and Emerging Results (ICSE-NIER), p. 29–32. ACM (2020)
59. Munassar, N.M.A., Govardhan, A.: A Comparison Between Five Models Of Software Engineering. International Journal of Computer Science Issues (IJCSI) **7**(5), 94 (2010)
60. Nielsen, M.A., Chuang, I.: Quantum Computation and Quantum Information (2002)
61. OASIS: Web Services Business Process Execution Language (WS-BPEL) Version 2.0. Organization for the Advancement of Structured Information Standards (2007)
62. OASIS: Topology and Orchestration Specification for Cloud Applications (TOSCA) Version 1.0. Organization for the Advancement of Structured Information Standards (2013)
63. OMG: Business Process Model and Notation (BPMN) Version 2.0. Object Management Group (2011)
64. Pérez-Castillo, R., Serrano, M.A., Piattini, M.: Software modernization to embrace quantum technology. Advances in Engineering Software **151**, 102933 (2021)
65. Pérez-Delgado, C.A., Perez-Gonzalez, H.G.: Towards a Quantum Software Modeling Language. In: Proceedings of the IEEE/ACM 42$^{nd}$ International Conference on Software Engineering Workshops, pp. 442–444 (2020)
66. Piattini, M., Peterssen, G., Pérez-Castillo, R.: Quantum Computing: A New Software Engineering Golden Age. ACM SIGSOFT Software Engineering Notes **45**(3), 12–14 (2020)
67. Piattini, M., Serrano, M., Perez-Castillo, R., Petersen, G., Hevia, J.L.: Toward a Quantum Software Engineering. IT Professional **23**(1), 62–66 (2021)
68. Pinter, S.S., Golani, M.: Discovering workflow models from activities' lifespans. Computers in Industry **53**(3), 283–296 (2004)
69. Preskill, J.: Quantum Computing in the NISQ era and beyond. Quantum **2**, 79 (2018)
70. Reed, M.D., et al.: Realization of three-qubit quantum error correction with superconducting circuits. Nature **482**(7385), 382–385 (2012)
71. Salm, M., Barzen, J., Breitenbücher, U., Leymann, F., Weder, B., Wild, K.: The NISQ Analyzer: Automating the Selection of Quantum Computers for Quantum Algorithms. In: Proceedings of the 14$^{th}$ Symposium and Summer School on Service-Oriented Computing (SummerSOC), pp. 66–85. Springer (2020)
72. Sete, E.A., Zeng, W.J., Rigetti, C.T.: A Functional Architecture for Scalable Quantum Computing. In: IEEE International Conference on Rebooting Computing, pp. 1–6 (2016)
73. Shor, P.W.: Polynomial-Time Algorithms for Prime Factorization and Discrete Logarithms on a Quantum Computer. SIAM Journal on Computing **26**(5), 1484–1509 (1997)
74. Simon, D.R.: On the power of quantum cryptography. In: 35$^{th}$ Annual Symposium on Foundations of Computer Science, pp. 116–123 (1994)
75. Sivarajah, S., Dilkes, S., Cowtan, A., Simmons, W., Edgington, A., Duncan, R.: t| ket>: A retargetable compiler for NISQ devices. Quantum Science and Technology (2020)
76. Sodhi, B., Kapur, R.: Quantum Computing Platforms: Assessing the Impact on Quality Attributes and SDLC Activities. arXiv:2104.14261 (2021)

77. Song, C., Cui, J., Wang, H., Hao, J., Feng, H., Li, Y.: Quantum computation with universal error mitigation on a superconducting quantum processor. Science advances **5**(9) (2019)
78. Suchara, M., Kubiatowicz, J., Faruque, A., Chong, F.T., Lai, C.Y., Paz, G.: QuRE: The Quantum Resource Estimator Toolbox. In: Proceedings of the 31st International Conference on Computer Design (ICCD), pp. 419–426. IEEE (2013)
79. Tannu, S.S., Qureshi, M.K.: Not all qubits are created equal: A case for variability-aware policies for nisq-era quantum computers. In: Proceedings of the 24th International Conference on Architectural Support for Programming Languages and Operating Systems, pp. 987–999 (2019)
80. Usaola, M.P.: Quantum Software Testing. In: Proceedings of the 1st International Workshop on the Quantum Software Engineering & Programming, pp. 57–63 (2020)
81. Vietz, D., et al.: On Decision Support for Quantum Application Developers: Categorization, Comparison, and Analysis of Existing Technologies. In: Proceedings of the 21th International Conference on Computational Science (ICCS), pp. 127–141. Springer (2021)
82. Wang, D., Higgott, O., Brierley, S.: Accelerated Variational Quantum Eigensolver. Physical review letters **122**(14), 140504 (2019)
83. Wang, S.A., Lu, C.Y., Tsai, I.M., Kuo, S.Y.: An XQDD-Based Verification Method for Quantum Circuits. IEICE transactions on fundamentals of electronics, communications and computer sciences **91**(2), 584–594 (2008)
84. Waters, B.R., Balfanz, D., Durfee, G., Smetters, D.K.: Building an Encrypted and Searchable Audit Log. In: NDSS, vol. 4, pp. 5–6. Citeseer (2004)
85. Weder, B., Barzen, J., Leymann, F., Salm, M.: Automated Quantum Hardware Selection for Quantum Workflows. Electronics **10**(8) (2021)
86. Weder, B., Barzen, J., Leymann, F., Salm, M., Vietz, D.: The Quantum Software Lifecycle. In: Proceedings of the 1st ACM SIGSOFT International Workshop on Architectures and Paradigms for Engineering Quantum Software (APEQS), pp. 2–9. ACM (2020)
87. Weder, B., Barzen, J., Leymann, F., Salm, M., Wild, K.: QProv: A Provenance System for Quantum Computing. IET Quantum Communication (2021)
88. Weder, B., Barzen, J., Leymann, F., Zimmermann, M.: Hybrid Quantum Applications Need Two Orchestrations in Superposition: A Software Architecture Perspective. In: Proceedings of the IEEE International Conference on Web Services (ICWS). IEEE (2021)
89. Weder, B., Breitenbücher, U., Képes, K., Leymann, F., Zimmermann, M.: Deployable Self-contained Workflow Models. In: Proceedings of the 8th European Conference on Service-Oriented and Cloud Computing (ESOCC), pp. 85–96. Springer (2020)
90. Weder, B., Breitenbücher, U., Leymann, F., Wild, K.: Integrating Quantum Computing into Workflow Modeling and Execution. In: Proceedings of the 13th IEEE/ACM International Conference on Utility and Cloud Computing (UCC), pp. 279–291. IEEE (2020)
91. Weigold, M., et al.: Data Encoding Patterns For Quantum Computing. In: Proceedings of the 27th Conference on Pattern Languages of Programs. The Hillside Group (2021)
92. Wettinger, J., Breitenbücher, U., Kopp, O., Leymann, F.: Streamlining DevOps automation for Cloud applications using TOSCA as standardized metamodel. Future Generation Computer Systems **56**, 317–332 (2016)
93. Wild, K., et al.: TOSCA4QC: Two Modeling Styles for TOSCA to Automate the Deployment and Orchestration of Quantum Applications. In: Proceedings of the 24th International Enterprise Distributed Object Computing Conference (EDOC), pp. 125–134. IEEE (2020)
94. Wu, Y., et al.: UML-Based Integration Testing for Component-Based Software. In: International Conference on COTS-Based Software Systems, pp. 251–260. Springer (2003)
95. Wurster, M., et al.: The Essential Deployment Metamodel: A Systematic Review of Deployment Automation Technologies. Software-Intensive Cyber-Physical Systems (2019)
96. Zapata: Orquestra (2021). URL https://www.zapatacomputing.com/orquestra
97. Zhao, J.: Quantum software engineering: Landscapes and horizons. arXiv:2007.07047 (2020)
98. Zimmermann, M., et al.: Towards Deployable Research Object Archives Based on TOSCA. In: Papers from the 12th Advanced Summer School on Service-Oriented Computing (SummerSoC), pp. 31–42. IBM Research Division (2018)

All links were last followed on June 15, 2021.